# Brain Responses During Robot-Error Observation

Dominik Welke [1 2 †], Joos Behncke [1 2 3 †], Marina Hader [2 4],
Robin Tibor Schirrmeister [1 2], Andreas Schönau [1 5], Boris Eßmann [1 5],
Oliver Müller [1 5], Wolfram Burgard [1 3], Tonio Ball [1 2 *]

[1] *BrainLinks–BrainTools, University of Freiburg, Germany.*
[2] *Translational Neurotechnology Lab, Epilepsy Center,
Medical Center –University of Freiburg, Faculty of Medicine,
University of Freiburg, Germany.*
[3] *Autonomous Intelligent Systems Group, Department of
Computer Science, University of Freiburg, Germany.*
[4] *Forschungszentrum Jülich, Germany.*
[5] *Department of Philosophy, University of Freiburg.*
[†] *These authors contributed equally.*

*\* Correspondence: Tonio Ball, BrainLinks-BrainTools, University of Freiburg,
c/o Translational Neurotechnology Lab, Engelbergerstr. 21, D-79106 Freiburg
E-mail: tonio.ball@uniklinik-freiburg.de*

**Abstract:** Brain-controlled robots are a promising new type of assistive device for severely impaired persons. Little is however known about how to optimize the interaction of humans and brain-controlled robots. Information about the human's perceived correctness of robot performance might provide a useful teaching signal for adaptive control algorithms and thus help enhancing robot control. Here, we studied whether watching robots perform erroneous vs. correct action elicits differential brain responses that can be decoded from single trials of electroencephalographic (EEG) recordings, and whether brain activity during human-robot interaction is modulated by the robot's visual similarity to a human.

To address these topics, we designed two experiments. In experiment I, participants watched a robot arm pour liquid into a cup. The robot performed the action either erroneously or correctly, i.e. it either spilled some liquid or not. In experiment II, participants observed two different types of robots, humanoid and non-humanoid, grabbing a ball. The robots either managed to grab the ball or not.

We recorded high-resolution EEG during the observation tasks in both experiments to train a Filter Bank Common Spatial Pattern (FBCSP) pipeline on the multivariate EEG signal and decode for the correctness of the observed action, and for the type of the observed robot. Our findings show that it was possible to decode both correctness and robot type for the majority of participants significantly, although often just slightly, above chance level. Our findings suggest that non-invasive recordings of brain responses elicited when observing robots indeed contain decodable information about the correctness of the robot's action and the type of observed robot. Our study also indicates that, given the, so far, relatively low decoding accuracies, either further improvements in non-invasive recording and analysis techniques or the utilization of intracranial measurements of neuronal activity will be necessary for practical applications.

## 1. INTRODUCTION

Autonomous technical systems are increasingly accessing our everyday life: The industry has been using robots for construction and assembly for years, autonomous cars are under development, and first robots especially designed for private users or social interaction (e.g., NAO (TM), Aldebaran Robotics, Paris, France or PARO Therapeutic Robot (TM), Intelligent System Co., Japan) already entered the open market. There is no reason to assume that this trend should lose momentum: Especially healthcare is a very promising field for robotic development with possible applications including robot-assisted surgery, motor analysis, rehabilitation, mental, cognitive and social therapy as well as robot-based patient monitoring systems.

Furthermore, robotic devices are among the key effectors in present and future applications of Brain-Computer-Interface (BCI) systems. BCIs are communication channels between a brain and a computer (Zander et al., 2007). Relying on learning algorithms, BCIs allow controlling the behavior of external devices, such as computers or exoskeletons (Bogue, 2009; Frisoli et al., 2012).

There has been more than a decade of research on human-robot interaction (see Goodrich and Schultz (2007) for a review), mostly in the fields of robotics and psychology, during which great importance has been assigned to the question of how to make interaction with robots most intuitive and "natural" for the human user (Dautenhahn et al., 2005). Literature on human-robot interaction identifies, among others, two important issues to be addressed in order to enhance a natural user experience: One question is how to enable robots to "read" human signals, both for control (Nickel and Stiefelhagen, 2007) and to detect errors in their own performance (Bartlett et al., 2003). The second point is to assess the influence of a robot's appearance and/or behavior on the user's cooperation and feelings towards the robot (Bartneck et al., 2009).

The goal of our study was to address both these issues from the perspective of neuroscience, which is so far only weakly represented in the research on human-robot interaction. On the one hand, we wanted to investigate which brain signals can be detected and thus be "read" by a robot (aided with machine learning techniques) to optimize robot behavior. On the other hand, we aimed to investigate the influence of the robot's visual similarity to a human on such error-related brain activity. For this purpose, we conducted two experiments, in which participants watched different kinds of robots perform correct and erroneous actions. In the following, we review electrophysiological studies on the observation of human and robot action and error-related brain responses relevant in this context. Then, we describe our hypotheses derived from the previous research and our experimental approach to test them.

*1.1 Brain Responses During Observation of Correct Human and Robot Actions*

One of the most striking findings in recent neuroscience was the discovery of mirror neurons: Rizzolatti et al. (1996) observed that certain neurons in the macaque brain fire both when the monkey performs an action and when it observes the same action performed by an experimenter. These cells were termed mirror neurons. The mirror neurons distributed across various brain regions together form what is called the mirror-neuron system (MNS; Rizzolatti and Craighero, 2004). Findings from neurophysiological and brain-imaging studies indicate that a MNS also exists in the human brain, possibly even spatially more extended than in monkeys, and that the MNS is reliably activated when humans observe other humans perform meaningful actions (Mukamel et al., 2010; Rizzolatti and Craighero, 2004).

Until now, there are very few published neurophysiological experiments on the perception of robotic action by a human observer. If it was processed similar to human movement, the human MNS should be involved during the observation of robot action. An electroencephalography (EEG) study by Oberman et al. (2007) suggests that the human MNS is indeed not selective for biological movement but can also be activated by robotic movement: Observation of a grasping action (target-directed and non-target-directed) performed by a robotic arm lead to suppression of mu-band activity (8-13 Hz) in left and right sensorimotor cortex (scalp positions C3 and C4), which has been linked to MNS activity. The study also found a significant hemispherical effect, as mu-band suppression was stronger on electrode C3 (left) than on C4 (right). Mu-band suppression also occurred when observing human motion. There was no significant difference in the strength of mu-band suppression during human vs. robot motion observation (Oberman et al., 2007).

*1.2 Error-Related Brain Responses*

Prior research found that human brain activity is modulated by both performed and observed erroneous action in other humans. When humans observe other humans committing errors or when they err themselves, their brains show a specific activation pattern in response to these errors (see Yeung et al. (2004) for a review). Bates et al. (2005) and Oberman et al. (2007) suggested that the MNS is involved in the error-observation-related brain responses.

With respect to the time domain of human EEG signals, there are well-documented event-related potential components (ERPs), which are linked to the processing of errors (mainly investigated in erroneous motor-execution): the error-related negativity (ERN), consisting of a negative deflection (Ne) and sometimes followed by a positive deflection (Pe). While the Pe seems to appear exclusively in conscious error processing (Nieuwenhuis et al., 2001), the Ne can be measured when participants do not report to have committed an error and a small negativity often even appears in correct trials (Vidal et al., 2003). Ne and Pe do not share the same scalp distribution: The Ne is maximal over fronto-central areas while the Pe is usually recorded with a parietal maximum (Falkenstein et al., 2000).

In the frequency domain of EEG signals, several studies demonstrated frequency-specific power modulations in response to erroneous action-execution in different motor tasks: Effects were mainly found in lower frequency bands, such as delta (1.5 - 3.5 Hz), theta (4 - 7 Hz) (Kolev et al., 2005; Yordanova et al., 2004), alpha (10 - 14 Hz) (Carp and Compton, 2009), and beta (15 - 30 Hz) bands (Vidal et al., 2003). Carp and Compton (2009) also suggested error-related spectral power changes in frequencies higher than 30 Hz. In a recent EEG study, our group found first evidence for modulations in the high-gamma range (50 - 150 Hz) related to erroneous execution of the Eriksen-Flanker motor task (Völker, 2015), a response inhibition test. Koelewijn et al. (2008) demonstrated an effect of the correctness of *observed* human actions on beta power-modulation over sensorimotor areas: This effect, however, was weaker in the observation settings than in a corresponding execution task.

In the context of BCIs for directly controlled prostheses, Milekovic et al. (2013) used electrocorticographic recordings obtained while participants where engaged in a simple videogame, for which they controlled a cursor with an analogue joystick. The experimenters were able to detect execution errors (i.e. when motor commands resulted in an unexpected movement) and outcome errors (i.e. when participants failed to reach the intended goal) from the neural activity in real-time significantly above chance level.

Iturrate et al. (2015) used EEG-measured error-related potentials in order to teach neuroprosthetics suitable behaviors in scenarios of varying complexity. In three experiments, participants were asked to monitor a device as it tried to reach a goal, that only the participant was aware of, and assess whether the actions of the device were incorrect or correct (i.e., whether the actions brought the device closer to the intended goal or further away). In experiment 1, participants observed a cursor on a screen as it moved either right or left in order to reach the target. In experiment 2, a simulated virtual robotic arm, that could perform four different actions (moving left, right, up or down) to reach the target, was displayed on a screen. For experiment 3, the simulated robotic arm was replaced by a real robotic arm. All experiments were divided into a training phase, during which the classifier that should detect the error-related potentials was built, and an online operation phase, during which the decoded information on correctness of the device's action was used as a reward for a reinforcement learning algorithm. Iturrate et al. (2015) were able to show that in 11 out of 12 participants, classification performance was significantly above chance level and that the user-controlled device reached the goal significantly more often as compared to a device following a random control policy. These findings demonstrate that error-related potentials are an adequate reward signal for reinforcement learning algorithms with the purpose of neuroprosthetics control.

Following the same line of research, Salazar-Gomez et al. (2017) investigated the role of EEG-measured error-related potentials for robot control during an object selection task in four conditions. In the online closed-loop condition, participants observed the robot perform binary object selection. If the EEG classifier detected an error-related potential, the robot's behavior was corrected, which in turn was immediately observed by the participant. In the offline closed-loop condition, the EEG classifier was trained using the data from all closed-loop trials of each participant. In the offline open-loop condition, participants observed the robot perform object selection correctly or incorrectly and no feedback was given to the robot. In the fourth condition, secondary errors, which occur in response to real-time misclassification (i.e., when the EEG signals are misclassified leading to incorrect robot behavior), were additionally considered in the online closed-loop condition. Performance in the online closed-loop condition was around chance level and on average above in the offline closed-loop and the offline open-loop condition. Interestingly, Salazar-Gomez et al. (2017) found that taking into consideration secondary errors improved performance significantly.

*1.3 Humanoid Robots*

To our knowledge, no study exists that directly compares the perception of humanoid robots to the perception of non-humanoid robots, with respect to the underlying brain responses. Many previous studies focused on how humanoid robots are perceived by humans with respect to facial features. DiSalvo et al. (2002) found that the presence of certain features in a robot's face, such as eyes, nose and mouth, the dimensions of the robot's head as well as the total number of facial features play a key role for the perceived humanness of a robot. Assuming the findings on facial features of robots can be transferred to general body features of robots, it appears likely that the higher the number of individual humanoid body features of a robot, the more humanoid it is perceived as a whole.

*1.4 Present Study*

Based upon the findings by Oberman et al. (2007) and van Schie et al. (2004), we hypothesized that watching a robot perform erroneous compared to correct action differentially modulates the observer's brain activity: Given that watching other humans performing erroneous actions triggers an automatic cognitive evaluation reflected in an error-related brain response (van Schie et al., 2004) and that observation of robot movement is processed similar to the observation of human movement (Oberman et al., 2007), we assumed that watching robots commit errors also elicits error-related brain responses. We propose that information about the perceived correctness of robot performance decoded from the EEG could provide a useful teaching signal for adaptive control algorithms in order to optimize robot control, in particular for so-called shared-control BCIs. Based on the results by DiSalvo et al. (2002), who were able to show that the number of humanlike features in a robot affects the perceived humanness of a robot, we were further interested whether this degree of perceived humanness would also be reflected in the brain responses that occur when observing a robot perform an action (both for actions where the robot commits an error and where it did not).

We designed two experiments to test these hypotheses. In experiment I, participants watched a robot arm pour liquid from a non-transparent container into a cup. The robot performed the action either incorrectly or correctly, i.e. it either spilled some liquid or not. In experiment II, a 2x2 factorial design was employed. Participants observed two different kinds of robots, a humanoid and a non-humanoid, grabbing a ball. Similar to experiment I, each of the robots was either successful at the action, i.e. managed to grab the ball, or not. Our approach was to decode the correctness of observed robot actions from the EEG signals recorded during the two different passive observation tasks and for experiment II, we also aimed to decode the type of the observed robot from the EEG signal.

Decoding was implemented using the common spatial pattern (CSP) approach for feature extraction, as applied to multi-channel EEG data by Müller-Gerking et al. in 1999 for decoding motor tasks (Müller-Gerking et al., 1999). Regularized linear discriminant analysis (rLDA) was used a classifier on these features. Since the the original studies, CSP has continuously been adapted and become a standard method in EEG classification tasks, especially for motor behavior or motor imagery (Blankertz et al., 2008). To validate the reliability of CSP decoding results, we assessed the spatial filters and corresponding activation patterns computed by the CSP algorithm.

As described above, previous studies have focused on brain responses during observation of correct and incorrect actions in humans but only few studies investigated the brain responses to watching robot action. To our knowledge, brain responses to perception of correct and incorrect robot performance and with different types of robots have never been assessed in previous experiments. Hence, the present study aimed to add to this field by investigating these aspects of robot observation.

## 2. EXPERIMENTAL PROCEDURES

*2.1 Experimental Paradigms*

*Experiment I: KUKA Pouring (see fig. 1 a)*

In both experiments, participants observed a set of short videos. The videos were presented repeatedly in randomized order. In experiment I, participants were shown videos of a robotic arm (LBR IIWA, KUKA Roboter GmbH, Augsburg, Germany) pouring orange juice from a non-transparent container into a cup. There were two classes of videos: The juice was either correctly poured into the cup, or incorrectly spilled over the table. Movement of the robotic arm and position of the cup were the same for all videos and conditions. Different outcomes were accomplished by varying the amount of juice in the container. The participants were thus unable to predict the outcome of the pouring action before it started. There were ten different video stimuli (five correct, five incorrect) of a 7.6-s length, with a frame rate of 30 frames per second (fps). The orange juice became first visible between 2.6 s and 3.23 s after the start of the video.

*Experiment II: NAO vs. NoHu (see fig. 1 b)*

Two different robots, either a small humanoid (NAO - Aldebaran Robotics, Paris, France) or a non-humanoid (custom-built, referred to as NoHu), approached a ball and tried to grab and lift it. In non-erroneous trials, the robots managed to grab and lift the ball and in erroneous trials, they failed to do so. There were 40 different video stimuli each of a 7-s duration, also with a frame rate of 30 fps (ten for each of the four conditions NAO - correct, NoHu - correct, NAO - incorrect, NoHu – incorrect; 5 of each set of 10 videos with the robot approaching from the left and 5 from the right, respectively). The initial position of the ball was invariant in all videos.

*General Paradigm (see fig. 1 c)*

Before the video stimuli were presented, participants fixed their gaze on a white fixation cross on gray background for baseline recording (3 s in experiment I, 2 s in experiment II). Then, the video was initiated (7.6 s in experiment I, 7 s in experiment II). In experiment II, 1 s of post-baseline activity was recorded to exclude preceding motor artifacts generated by answering the attention task (see fig. 1). Up to 10 s of time between trials followed, allowing the participants to move, blink, swallow and answer a simple attention control question (Action correct? – Yes/No). The fixation cross was shown on top of the video display in the area of the main events of interest: For experiment I, this was the area of the cup in the lower right part of the screen and for experiment II, it was the initial place of the ball in the center of the screen. The control question was displayed at the same position. It was answered by pressing a key on a keypad-controller. Respective keys for answers "yes" and "no" were switched every 40 trials. After self-paced answering of the control question, the subsequent trial was initiated. The experiment was conducted in sessions of 30 trials in experiment I and of 40 trials in experiment II, respectively. Trigger pulses containing an unambiguous ID were generated with the onset of video presentation and recorded via the EEG amplifiers. An additional optical trigger for post-hoc reconstruction in combination with a photo diode was embedded in the video.

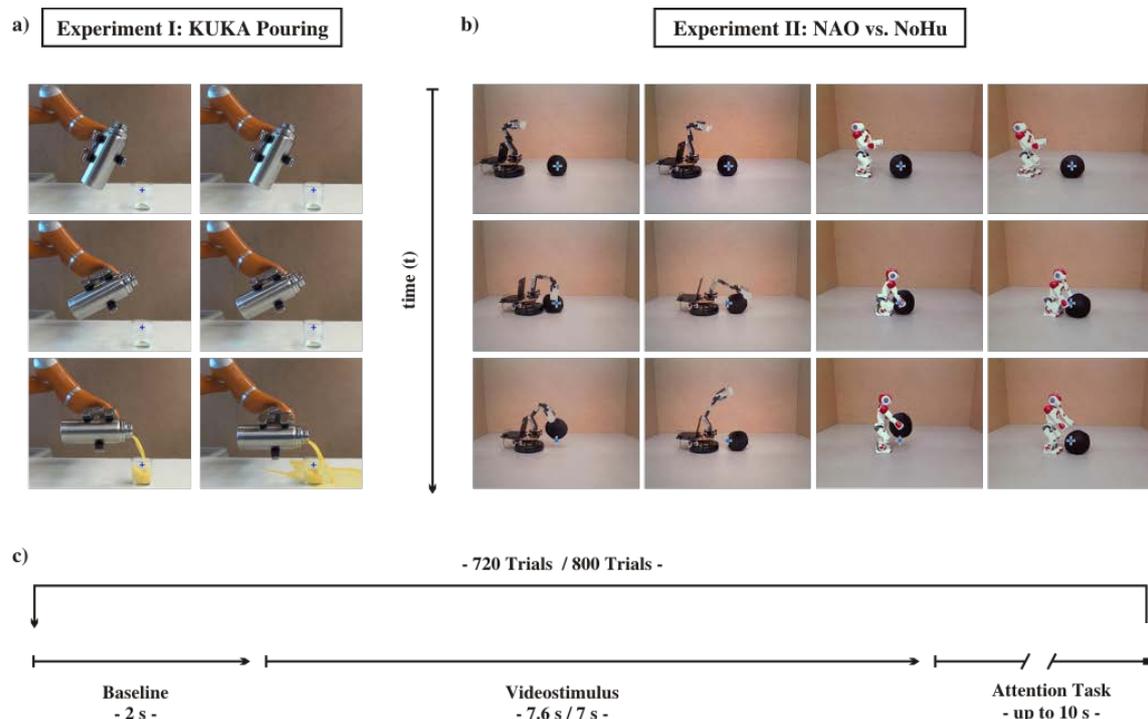

Figure 1. Experimental paradigms: video stimuli in exp. I (a) and exp. II (b), and schematic procedure of the observation tasks (c).

In experiment I, we recorded at least 720 trials per participant (360 trials per condition, 72 trials per video). In participants 1 to 3, the number of trials per class was unbalanced (60% correct to 40% incorrect). Therefore, a weighing-mechanism was included in the analysis (see below). In experiment II, we recorded at least 800 trials per participant (200 trials per condition, 20 trials per video). For each participant, the experiment including preparation and pauses lasted about between 5 to 6 hours for experiment I and between 5 to 7 hours for experiment II.

*2.2 Participants*

The participants were included in the study upon their informed written consent. All were healthy adults, either students or PhD students (22 to 31 years old). In experiment I, 6 participants (6 male) took part; one participant was excluded due to insufficient number of trials. In experiment II, a total of 12 participants took part (6 female); one participant was excluded due to insufficient number of trials. The study was approved by the Ethics Committee of the University Medical Center Freiburg.

*2.3 Data Acquisition*

Experiments were conducted in an electromagnetically shielded cabin ("mrShield" – CFW Trading Ltd., Heiden, Switzerland). All electric devices in the cabin were powered by DC batteries. Information between inside and outside of the cabin was exchanged only via fiber-optic cables. We used high-precision EEG amplifiers with a 24-bit digital resolution and low noise (NeurOne - Mega Electronics Ltd., Kuopio, Finland) to record EEG from 128 scalp positions according to the "five percent" electrode-layout (Waveguard 128 - ANT Neuro, Netherlands). The gel-filled electrodes were prepared to reach impedances below 5 kΩ if possible. Sampling rate was 5 kHz; electrode Cz was used as reference, the ground was located between AFz- and Fz-position.

We also recorded electrooculograms (EOGs), electrocardiograms (ECGs) and electromyograms (EMGs) of arms and legs of the participants and additionally used an infra-red eye-tracker to monitor eye movements (EyeLink 1000+ - SR Research Ltd., Canada). Eye-tracking data and EOG was used to inspect additionally whether participants looked at the stimuli; EMG to verify that participants remained still; ECG recordings were not used in the present analyses.

*2.4 Data Analysis*

*Signal Pre-Processing*

Data was down-sampled to 500 Hz, and then high-pass filtered with a cut-off frequency of 0.5 Hz using a stable 4th order Butterworth filter. Noisy channels were determined by visual inspection first and post hoc by using an automatic cleaning algorithm optimized to detect muscle-artifacts based upon the variance of the signal (BBCI-Toolbox, Blankertz et al., 2010). To identify noisy trials, the data were analyzed in intervals corresponding to decoding intervals plus a preceding 500 ms and the trials were rejected if the difference between the maximum and minimum value exceeded 600 µV. Rejected trials were only excluded from the training sets but kept in the test sets of the cross-validation (see below). Then, common-average re-referencing was performed and trials were cut according to the decoding intervals described below.

*Decoding*

We implemented a Filter Bank Common Spatial Pattern (FBCSP) algorithm following Ang et al. (2012). The data was bandpass-filtered in 35 non-overlaying frequency bands between 0.5 Hz and 144 Hz. Between 0.5 Hz and 30 Hz, a filter with a bandwidth of 2 Hz was applied and between 30 Hz and 144 Hz – with a bandwidth of 6 Hz, since band power modulations in low frequencies typically occur in narrower bands than in high frequencies (Buzsáki and Draguhn, 2004). CSP analysis was then performed on each of these frequency bands in a 10-fold cross-validation: The feature selection was set to choose the first 3 and the last 3 filters ordered according to their eigenvalues, i.e. the most discriminative six filters (see Blankertz et al. (2008) for more details on this heuristic), to maximize between-class variance in the training set. These spatial filters were then applied on the trial data. The logarithm of the variance of the resulting signal was used as features. Then, as a first step, rLDA classifiers (Blankertz et al., 2010; Friedman, 1989) were trained on the training features and evaluated on the test features, which resulted in frequency-resolved decoding accuracies. In a second step, the stored features from either all frequency-bands or two different subsets (all frequency bands below 20 Hz, and all frequency bands above 60 Hz) were taken together to train FBCSP classifiers (in a 10-fold cross-validation analogous to the above). To account for unbalanced numbers of trials in the different classes, the mean over decoding accuracies per class was used instead of the overall decoding accuracy.

For experiment I, binary FBCSP for the classes correct vs. incorrect was performed in the decoding intervals

0 - 7600 ms (full interval) and 3300 - 7500 ms (late interval) relative to start of the video display. In addition, the interval from -500 – 3000 ms relative to the point in time when the liquid first became visible was extracted. This interval (intermediate interval) differed depending on the video displayed and accounted for the fact that the frame where liquid first become visible varied among the stimuli; liquid in incorrect trials appeared between 400 to 600 ms earlier than in correct trials.

For experiment II, binary FBCSP for the classes NAO vs. NoHu as well as for the classes correct vs. incorrect was performed in the intervals 0 - 7000 ms (full interval), 5100 - 6900 ms (late interval) and 4000 - 7000 ms (intermediate interval), relative to the start of the video display, to cover the different phases of the stimuli.

*2.5 Statistics*

P-values for FBCSP decoding accuracies for each participant were estimated by a randomization test as follows: For the given participant, classifications for all trials were randomly sampled from a uniform distribution over the original predictions for all trials and the resulting decoding accuracy was computed. This process was repeated 100000 times. The p-value was then estimated as the fraction of these repetitions that resulted in decoding accuracies at least as large as the accuracy resulting from the actual predictions by FBCSP.

## 3. RESULTS

Fig. 2 exemplarily shows both filter and corresponding activation patterns calculated for CSP decoding in experiment I (fig. 2 a), experiment II - error condition (fig. 2 b) and experiment II - robot condition (fig. 2 c). For visualization, we used the filters of participant 1 (experiment I) and of participant 2 (experiment II) which reached the highest decoding accuracies among all frequency bands below 20 Hz. The filters showed ordered spatial patterns pointing to (multiple) bipolar generators.

Fig. 3 shows mean CSP decoding accuracies in experiment I of all 35 frequency bands in the range between 0.5 - 144 Hz for participants 1 to 5 (decoding-interval 3300 - 7500 ms relative to video stimulus onset). Decoding accuracies were mainly above chance level (50%), in participants 1 to 4, accuracies were generally higher for frequency ranges below 20 Hz. This trend with respect to frequency ranges was also found in the other decoding intervals and (weaker) in experiment II (not shown). Participant 1 also showed above-average decoding accuracies in frequency bands beyond 60 Hz. Maximal decoding accuracies reached up to around 75% in participants 1 and 5. Fig. 4 a) compares decoding accuracies for FBCSP for different frequency ranges (broadband: 0.5 - 144 Hz, low frequencies: 0.5 - 19 Hz and high gamma: 61 - 144 Hz) for participants 1 to 5 (experiment I). Fig. 4 b) and c) compare FBCSP classifiers over the different frequency ranges for participants 1 to 11 (experiment II). The presented results originate from the decoding intervals which yielded the highest mean decoding accuracies: For experiment I, 3300 - 7500 ms

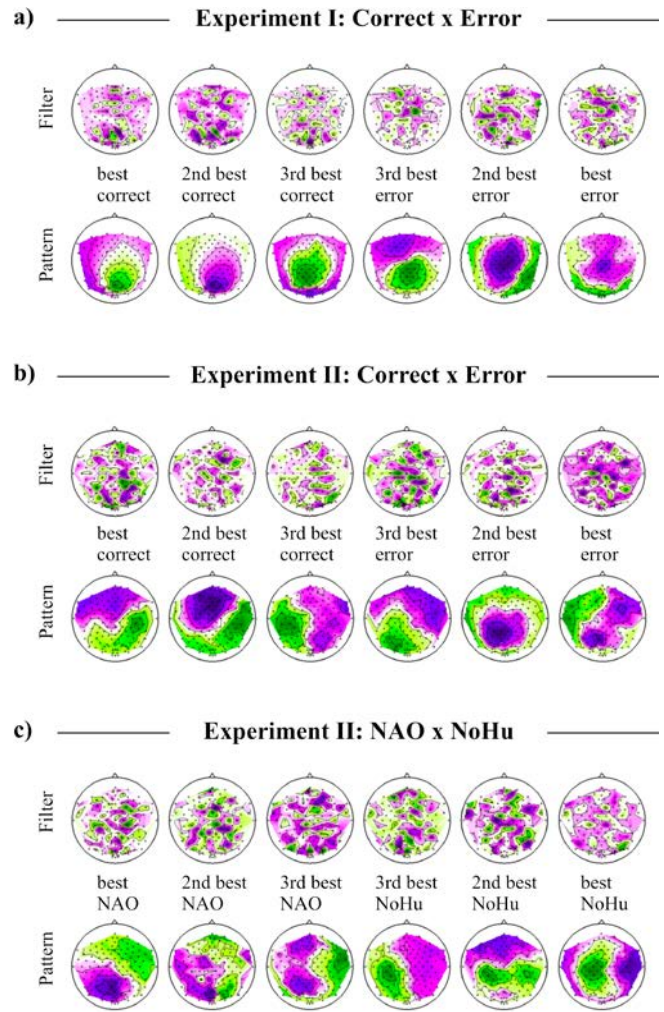

Figure 2. Exemplary CSP-filters and activation patterns for the error condition in experiment I (a), the error condition in experiment II (b), and the robot-type condition in experiment II (c).

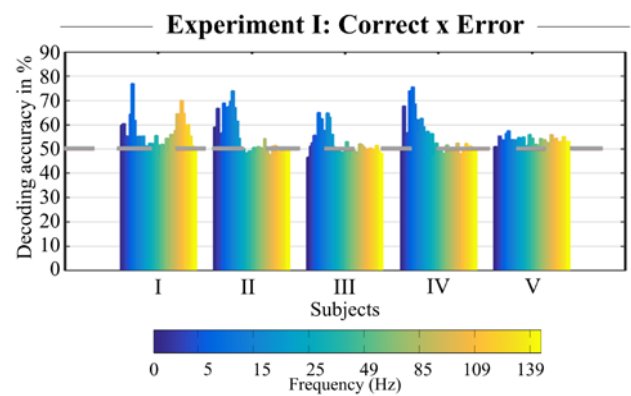

Figure 3. Frequency-resolved CSP-decoding results in experiment I.

(fig. 4 a), for experiment II (error condition) 4000 - 7000 ms (fig. 4 b) and for experiment II (robot condition) 0 - 7000 ms (fig. 4 c). Mean decoding accuracies over all participants for the different frequency ranges and decoding intervals can be found in tables 1 and 2. Table 1 contains decoding accuracies extracted from the best decoding interval (experiment I - error: late; experiment II - error: intermediate; experiment II - robot: full) whereas table 2 shows the decoding results of FBCSP for frequencies in the range of 0.5 – 19 Hz.

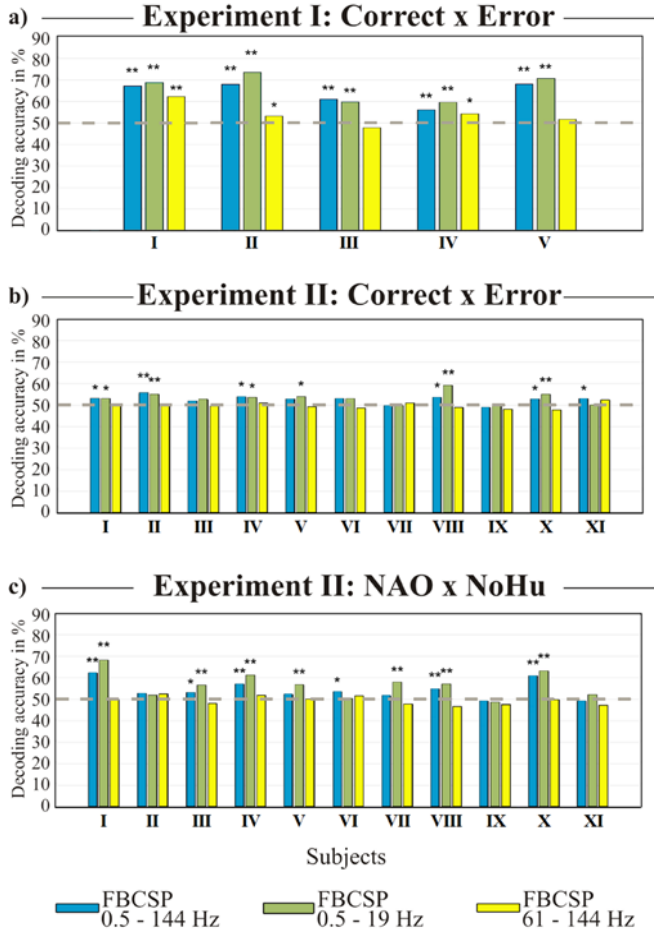

Figure 4. FBCSP decoding results for experiment I (a, interval: 3300 – 7500 ms), for the error condition of experiment II (b, interval: 4000 – 7000 ms) and robot condition of experiment II (c, interval: 0 – 7000 ms). Significance is indicated by asterisks: * p < 0.05, ** p < 0.01.

Table 1. FBCSP decoding results of the different classifiers

|  | Mean decoding accuracies and standard deviation | | |
|---|---|---|---|
|  | *0.5–144 Hz* | *< 20 Hz* | *> 60 Hz* |
| **Exp. I error** | 0.602 (0.053) | 0.621 (0.057) | 0.540 (0.053) |
| **Exp. II error** | 0.526 (0.019) | 0.532 (0.027) | 0.496 (0.014) |
| **Exp. II robot** | 0.542 (0.043) | 0.567 (0.058) | 0.494 (0.020) |

Table 2. FBCSP decoding results on the different time intervals

|  | Mean decoding accuracies and standard deviation | | |
|---|---|---|---|
|  | *full interval* | *late interval* | *intermediate interval* |
| **Exp. I error** | 0.552 (0.036) | 0.596 (0.075) | 0.580 (0.053) |
| **Exp. II error** | 0.511 (0.032) | 0.530 (0.025) | 0.532 (0.027) |
| **Exp. II robot** | 0.567 (0.058) | 0.534 (0.032) | 0.564 (0.043) |

## 4. DISCUSSION

Our findings show that both conditions investigated in the present study were decodable from the recorded EEG signals: observation of erroneous vs. correct robot actions in experiment I and, although at very low accuracies and not in all participants, also in experiment II, as well as humanoid vs. non-humanoid robot type in experiment II. FBCSP decoding accuracies varied in a range from around chance level to around 70% (fig. 4, tab. 1). The fact that erroneous vs. correct robot action can be decoded from human brain activity is line with prior findings by Iturrate et al. (2015). Taken together, our results indicate that observation of erroneous vs. correct robot action and observed robot type are encoded in human brain activity and that related brain signals can be detected, at least in some circumstances, in the non-invasively recorded EEG.

In contrast to promisingly high error decoding accuracies in experiment I, the accuracies for the broadband and low-frequency components of the error condition of experiment II were mostly little above chance level (fig. 4). This renders a generalizing of the results difficult and suggests that an unknown factor may have played a role. For example, one possible explanation could be that errors in experiment I were indirectly decoded by differences in the visual properties of the stimuli during correct and incorrect trials, which were more prominent in experiment I. However, differences in visual properties were also present in experiment II. As a matter of fact, they are an inevitable consequence of errors in everyday settings as investigated in the present study. Another explanation could be that different error types may elicit different affective responses, which in this case were possibly more pronounced in experiment I (liquid spilling) than in experiment II (unsuccessful ball grabbing). Follow-up studies will be necessary to further investigate these observations and generally the factors modulating robot-error recognition by humans as well as the underlying neurophysiology.

Our results indicate an effect of the EEG frequencies used for decoding. FBCSP taking into account only frequencies below 20 Hz yielded generally higher decoding accuracies than broadband or high-frequency components (tab. 1). This is in line with previous studies which suggested an involvement of the MNS and motor system, manifested in

mu- and beta-band modulations (Koelewijn et al., 2008; Oberman et al., 2007), as well as frequency power modulations in response to erroneous action execution mainly found in lower frequency bands such as delta, theta, alpha and beta (Carp and Compton, 2009; Kolev et al., 2005; Vidal et al., 2003; Yordanova et al., 2004). High frequency-components in the gamma range were previously also proposed to be related to error processing (Carp and Compton, 2009; Völker, 2015). Yet, in our experiment, FBCSP using frequency components of the high gamma range (above 60 Hz) yielded comparatively low decoding accuracies. This effect, however, seemed to be participant-dependent (see fig. 3), reminiscent of inter-participant variability of movement-related high gamma EEG responses as they have previously been observed (Ball et al., 2008; Demandt et al., 2012). Further investigations are necessary to elucidate the role of different EEG frequency bands, including gamma, and of their dynamics in the context of robot-error observation.

Our results also indicate a possible effect of the time intervals used for decoding. Decoding for correctness in both experiments, intervals starting after the stimulus condition had become evident to the observer (late and intermediate intervals) appeared to yield the best results (tab. 2). This suggests that the timespan preceding the error, which was designed to contain minimal visual differences, was indeed uninformative for decoding. In contrast, decoding for the robot type in experiment II, the full video interval yielded the best overall decoding results, while the shorter intervals resulted in lower accuracies, consistent with the fact that the robot type difference was present throughout the trials.

We investigated robots during naturalistic tasks, namely liquid pouring and object grabbing. These tasks were designed to approximate application scenarios of autonomous or semi-autonomous robots under high-level control or surveillance via brain-computer interfacing. An important consequence of these naturalistic conditions is that the exact time point of error events is less clearly defined than in other experimental paradigms that have previously been used to elicit error-related brain responses (e.g., in forced choice visual discrimination tasks, such as the Erikson Flanker task) or to investigate low-level BCI control. Some of these paradigms are able to yield much higher decoding accuracies than we observed here (up to 91% for single-trial classification; see Parra et al., 2003). Yet, with respect to future prospects of robotics and shared-control BCI applications, unpredictable, asynchronous errors are an important, complementary topic. Another, related perspective would be the detection and evaluation of action consequences that are not "objectively" right or wrong, but rather depend on the intention of the user (for example, picking up an apple vs. picking up a piece of chocolate).

The achieved accuracies in all experiments are not as high as would be required for improving BCI-applications. Chavarriaga et al. (2014) suggest 80% accuracy as a benchmark for decoding of error-related potentials that has been shown to be sufficient to improve information transfer rate in most BMI applications. Improvement of decoding accuracy in our experimental paradigm could be reached by resorting to other types of electrophysiological recordings: Even though we used a highly optimized EEG setup, intracranial recordings can be expected to provide more reliable error-related signals if recorded from informative brain areas. Despite the low accuracies reached here, non-invasive studies as in the present study combined with source localization approaches may be useful for guiding such studies and selecting the most promising target areas involved in the perception and cognitive evaluation of robot action.

The future will bring many major developments in the fields of robotics and shared-control BCIs: With growing relevance of robots in everyday life, new types of interaction between humans and machines will evolve. The current practice of surveying robots (e.g., Dautenhahn et al., 2005) could profit by elaborated, empirically supported theories. In this context, it may be fruitful to join efforts across scientific disciplines and, for example, include concepts and theories from philosophical action theory into empirical studies, as previously proposed by Thinnes-Elker et al. (2012). Action theory conceptualizes human agency by analyzing agency-related phenomena like intention and planning (for an overview see O'Connor and Sandis, 2011). A central topic of the emerging interdisciplinary field of "action science" (Prinz et al., 2013) is the integration of philosophical concepts with related empirical findings, e.g., from the field of neuroscience (cf. Butterfill and Sinigaglia, 2014; Haggard, 2017; Pacherie, 2008), into a comprehensive theory. In the past years, so-called "shared agency" among cooperating human agents (Bratman, 2013; Pacherie, 2011; Velleman, 1997; Vesper et al., 2010) became a focus of interdisciplinary studies in this area (Wenke et al., 2011) and very recent experimental work investigated shared agency during human-robot cooperation (e.g., Fiore et al., 2016; Hinds et al., 2004; Stubbs et al., 2007) as well as the agentive properties of robots in general (Khamassi et al., 2016). Such collaboration across disciplines has already proven successful in the field of action theory, where the application of concepts from philosophy to the field of intelligent systems has led to the development of belief-desire-intention architectures (Georgeff et al., 1999)

## 5. CONCLUSION

The findings presented in the present study indicate that it is possible to decode the correctness of at least some kinds of observed robotic actions as well as the type of observed robot from non-invasively recorded human EEG. These findings add to relevant topics in the research on human-robot interaction, such as enabling robotic systems to "read" human signals or the influence of a robot's appearance and/or behavior on the user's perception of the robot.

Accessing error recognition in robot performance might be helpful for EEG-based BCIs; our observation tasks were designed to approximate future application scenarios of autonomous or semi-autonomous robots under high-level control or surveillance via brain-computer interfacing (self-feeding, go-and-fetch tasks). There are several perspectives for follow-up investigations which derive from the present study, and which could be addressed with similar methods as we have employed. Given that the achieved accuracies are likely not yet sufficient for practical applications, it would be helpful if alternative machine learning approaches such as artificial neural networks reached higher decoding accuracies. Another question to be addressed in the future would be which kind of robot errors are generally suitable for decoding of the user's perceived correctness and how they differ from non-decodable errors. Closely related to this, would be the investigation of how visual, affective, and movement-related brain systems are involved in the generation of the differential responses to robot action.

## ACKNOWLEDGEMENTS

This work was supported by the DFG Excellence Cluster BrainLinks-BrainTools (EXC1086) and the BMI-Bot Project by the Baden-Württemberg-Stiftung.

## REFERENCES

Ang, K.K., Chin, Z.Y., Wang, C., Guan, C., Zhang, H., 2012. Filter Bank Common Spatial Pattern Algorithm on BCI Competition IV Datasets 2a and 2b. *Front. Neurosci.* 6. doi:10.3389/fnins.2012.00039

Ball, T., Demandt, E., Mutschler, I., Neitzel, E., Mehring, C., Vogt, K., Aertsen, A., Schulze-Bonhage, A., 2008. Movement related activity in the high gamma range of the human EEG. *NeuroImage* 41, 302–310. doi:10.1016/j.neuroimage.2008.02.032

Bartlett, M.S., Littlewort, G., Fasel, I., Chenu, J., Ishiguro, H., Movellan, J.R., 2003. Towards social robots: Automatic evaluation of human-robot interaction by face detection and expression classification, *Advances in.* MIT Press, p. 2003.

Bartneck, C., Kulić, D., Croft, E., Zoghbi, S., 2009. Measurement Instruments for the Anthropomorphism, Animacy, Likeability, Perceived Intelligence, and Perceived Safety of Robots. *Int. J. Soc. Robot.* 1, 71–81. doi:10.1007/s12369-008-0001-3

Bates, A.T., Patel, T.P., Liddle, P.F., 2005. External Behavior Monitoring Mirrors Internal Behavior Monitoring. *J. Psychophysiol.* 19, 281–288. doi:10.1027/0269-8803.19.4.281

Blankertz, B., Tangermann, M., Vidaurre, C., Fazli, S., Sannelli, C., Haufe, S., Maeder, C., Ramsey, L., Sturm, I., Curio, G., Müller, K.-R., 2010. The Berlin Brain–Computer Interface: Non-Medical Uses of BCI Technology. *Front. Neurosci.* 4. doi:10.3389/fnins.2010.00198

Blankertz, B., Tomioka, R., Lemm, S., Kawanabe, M., Muller, K., 2008. Optimizing Spatial filters for Robust EEG Single-Trial Analysis. *IEEE Signal Process. Mag.* 25, 41–56. doi:10.1109/MSP.2008.4408441

Bogue, R., 2009. Exoskeletons and robotic prosthetics: a review of recent developments. *Ind. Robot Int. J.* 36, 421–427. doi:10.1108/01439910910980141

Bratman, M.E., 2013. *Shared agency: A planning theory of acting together.* Oxford University Press.

Butterfill, S.A., Sinigaglia, C., 2014. Intention and Motor Representation in Purposive Action: INTENTION AND MOTOR REPRESENTATION IN PURPOSIVE ACTION. *Philos. Phenomenol. Res.* 88, 119–145. doi:10.1111/j.1933-1592.2012.00604.x

Buzsáki, G., Draguhn, A., 2004. Neuronal Oscillations in Cortical Networks. *Science* 304, 1926–1929. doi:10.1126/science.1099745

Carp, J., Compton, R.J., 2009. Alpha power is influenced by performance errors. *Psychophysiology* 46, 336–343. doi:10.1111/j.1469-8986.2008.00773.x

Chavarriaga, R., Sobolewski, A., Millán, J. del R., 2014. Errare machinale est: the use of error-related potentials in brain-machine interfaces. *Front. Neurosci.* 8. doi:10.3389/fnins.2014.00208

Dautenhahn, K., Woods, S., Kaouri, C., Walters, M.L., Koay, K.L., Werry, I., 2005. What is a robot companion - friend, assistant or butler?, in: 2005 IEEE/RSJ International Conference on Intelligent Robots and Systems. Presented at the 2005 *IEEE/RSJ International Conference on Intelligent Robots and Systems*, pp. 1192–1197. doi:10.1109/IROS.2005.1545189

Demandt, E., Mehring, C., Vogt, K., Schulze-Bonhage, A., Aertsen, A., Ball, T., 2012. Reaching Movement Onset- and End-Related Characteristics of EEG Spectral Power Modulations. *Front. Neurosci.* 6. doi:10.3389/fnins.2012.00065

DiSalvo, C.F., Gemperle, F., Forlizzi, J., Kiesler, S., 2002. All Robots Are Not Created Equal: The Design and Perception of Humanoid Robot Heads, in: *Proceedings of the 4th Conference on Designing Interactive Systems: Processes, Practices, Methods, and Techniques*, DIS '02. ACM, New York, NY, USA, pp. 321–326. doi:10.1145/778712.778756

Falkenstein, M., Hoormann, J., Christ, S., Hohnsbein, J., 2000. ERP components on reaction errors and their functional significance: a tutorial. *Biol. Psychol.* 51, 87–107. doi:10.1016/S0301-0511(99)00031-9

Fiore, M., Clodic, A., Alami, R., 2016. On Planning and Task Achievement Modalities for Human-Robot Collaboration, in: Hsieh, M.A., Khatib, O., Kumar, V. (Eds.), *Experimental Robotics.* Springer International Publishing, Cham, pp. 293–306. doi:10.1007/978-3-319-23778-7_20

Friedman, J.H., 1989. Regularized Discriminant Analysis. *J. Am. Stat. Assoc.* 84, 165. doi:10.2307/2289860

Frisoli, A., Loconsole, C., Leonardis, D., Banno, F., Barsotti, M., Chisari, C., Bergamasco, M., 2012. A New Gaze-BCI-Driven Control of an Upper Limb Exoskeleton for Rehabilitation in Real-World Tasks. *IEEE Trans. Syst. Man Cybern. Part C Appl. Rev.* 42, 1169–1179. doi:10.1109/TSMCC.2012.2226444

Georgeff, M., Pell, B., Pollack, M., Tambe, M., Wooldridge, M., 1999. The Belief-Desire-Intention

Model of Agency, in: Müller, J.P., Rao, A.S., Singh, M.P. (Eds.), *Intelligent Agents V: Agents Theories, Architectures, and Languages*. Springer Berlin Heidelberg, Berlin, Heidelberg, pp. 1–10. doi:10.1007/3-540-49057-4_1

Goodrich, M., Schultz, A., 2007. Human-robot interaction: a survey. *Found. Trends Hum.-Comput. Interact.* 1, 203–275. doi:10.1561/1100000005

Haggard, P., 2017. Sense of agency in the human brain. *Nat. Rev. Neurosci.* 18, 196–207. doi:10.1038/nrn.2017.14

Hinds, P.J., Roberts, T.L., Jones, H., 2004. Whose job is it anyway? A study of human-robot interaction in a collaborative task. *Hum.-Comput. Interact.* 19, 151–181.

Iturrate, I., Chavarriaga, R., Montesano, L., Minguez, J., Millán, J.D.R., 2015. Teaching brain-machine interfaces as an alternative paradigm to neuroprosthetics control. *Scientific reports*, 5.

Khamassi, M., Girard, B., Clodic, A., Devin, S., Renaudo, E., Pacherie, E., Alami, R., Chatila, R., 2016. Integration of Action, Joint Action and Learning in Robot Cognitive Architectures. *Intellectica- Rev. L'Association Pour Rech. Sur Sci. Cogn. ARCo* 2016, 169–203.

Koelewijn, T., van Schie, H.T., Bekkering, H., Oostenveld, R., Jensen, O., 2008. Motor-cortical beta oscillations are modulated by correctness of observed action. *NeuroImage* 40, 767–775. doi:10.1016/j.neuroimage.2007.12.018

Kolev, V., Falkenstein, M., Yordanova, J., 2005. Aging and Error Processing: Time-Frequency Analysis of Error-Related Potentials. *J. Psychophysiol.* 19, 289–297. doi:10.1027/0269-8803.19.4.289

Mehring, C., Nawrot, M.P., de Oliveira, S.C., Vaadia, E., Schulze-Bonhage, A., Aertsen, A., Ball, T., 2004. Comparing information about arm movement direction in single channels of local and epicortical field potentials from monkey and human motor cortex. *J. Physiol.-Paris* 98, 498–506. doi:10.1016/j.jphysparis.2005.09.016

Milekovic, T., Ball, T., Schulze-Bonhage, A., Aertsen, A., Mehring, C., 2013. Detection of error related neuronal responses recorded by electrocorticography in humans during continuous movements. *PloS One* 8, e55235. doi:10.1371/journal.pone.0055235

Milekovic, T., Fischer, J., Pistohl, T., Ruescher, J., Schulze-Bonhage, A., Ad Aertsen, Rickert, J., Ball, T., Mehring, C., 2012. An online brain–machine interface using decoding of movement direction from the human electrocorticogram. *J. Neural Eng.* 9, 046003. doi:10.1088/1741-2560/9/4/046003

Mukamel, R., Ekstrom, A.D., Kaplan, J., Iacoboni, M., Fried, I., 2010. Single-Neuron Responses in Humans during Execution and Observation of Actions. *Curr. Biol.* 20, 750–756. doi:10.1016/j.cub.2010.02.045

Müller-Gerking, J., Pfurtscheller, G., Flyvbjerg, H., 1999. Designing optimal spatial filters for single-trial EEG classification in a movement task. *Clin. Neurophysiol.* 110, 787–798. doi:10.1016/S1388-2457(98)00038-8

Nickel, K., Stiefelhagen, R., 2007. Visual recognition of pointing gestures for human–robot interaction. *Image Vis. Comput.* 25, 1875–1884. doi:10.1016/j.imavis.2005.12.020

Nieuwenhuis, S., Ridderinkhof, K.R., Blom, J., Band, G.P.H., Kok, A., 2001. Error-related brain potentials are differentially related to awareness of response errors: Evidence from an antisaccade task. *Psychophysiology* 38, 752–760. doi:10.1111/1469-8986.3850752

Oberman, L.M., McCleery, J.P., Ramachandran, V.S., Pineda, J.A., 2007. EEG evidence for mirror neuron activity during the observation of human and robot actions: Toward an analysis of the human qualities of interactive robots. *Neurocomputing,* Selected papers from the 3rd International Conference on Development and Learning (ICDL 2004) Time series prediction competition: the CATS benchmark3rd International Conference on Development and Learning 70, 2194–2203. doi:10.1016/j.neucom.2006.02.024

O'Connor, T., Sandis, C., 2011. *A Companion to the Philosophy of Action*. John Wiley & Sons.

Pacherie, E., 2011. Framing Joint Action. *Rev. Philos. Psychol.* 2, 173–192. doi:10.1007/s13164-011-0052-5

Pacherie, E., 2008. The phenomenology of action: A conceptual framework. *Cognition* 107, 179–217. doi:10.1016/j.cognition.2007.09.003

Parra, L.C., Spence, C.D., Gerson, A.D., Sajda, P., 2003. Response error correction-a demonstration of improved human-machine performance using real-time eeg monitoring. *IEEE Trans. Neural Syst. Rehabil. Eng.* 11, 173–177. doi:10.1109/TNSRE.2003.814446

Pistohl, T., Schulze-Bonhage, A., Aertsen, A., Mehring, C., Ball, T., 2012. Decoding natural grasp types from human ECoG. *NeuroImage* 59, 248–260. doi:10.1016/j.neuroimage.2011.06.084

Prinz, W., Beisert, M., Herwig, A. (Eds.), 2013. *Action science: foundations of an emerging discipline*. MIT Press, Cambridge, Mass.

Rizzolatti, G., Craighero, L., 2004. The mirror-neuron system. *Annu. Rev. Neurosci.* 27, 169–192. doi:10.1146/annurev.neuro.27.070203.144230

Rizzolatti, G., Fadiga, L., Gallese, V., Fogassi, L., 1996. Premotor cortex and the recognition of motor actions. *Cogn. Brain Res.*, Mental representations of motor acts 3, 131–141. doi:10.1016/0926-6410(95)00038-0

Salazar-Gomez, A.F., Del Preto, J., Gil, S., Guenther, F., Rus, D., 2017. Correcting Robot Mistakes in Real Time Using EEG Signals. Presented at the 2017 *IEEE International Conference on Robotics and Automation (ICRA)*.

Stubbs, K., Hinds, P.J., Wettergreen, D., 2007. Autonomy and Common Ground in Human-Robot Interaction:

A Field Study. *IEEE Intell. Syst.* 22, 42–50. doi:10.1109/MIS.2007.21

Thinnes-Elker, F., Iljina, O., Apostolides, J.K., Kraemer, F., Schulze-Bonhage, A., Aertsen, A., Ball, T., 2012. Intention Concepts and Brain-Machine Interfacing. *Front. Psychol.* 3. doi:10.3389/fpsyg.2012.00455

van Schie, H.T., Mars, R.B., Coles, M.G.H., Bekkering, H., 2004. Modulation of activity in medial frontal and motor cortices during error observation. *Nat. Neurosci.* 7, 549–554. doi:10.1038/nn1239

Velleman, J.D., 1997. How To Share An Intention. *Philos. Phenomenol. Res.* 57, 29. doi:10.2307/2953776

Vesper, C., Butterfill, S., Knoblich, G., Sebanz, N., 2010. A minimal architecture for joint action. *Neural Netw.* 23, 998–1003. doi:10.1016/j.neunet.2010.06.002

Vidal, F., Burle, B., Bonnet, M., Grapperon, J., Hasbrocq, T., 2003. Error negativity on correct trials: a reexamination of available data. *Biol. Psychol.* 64, 265–282. doi:10.1016/S0301-0511(03)00097-8

Völker, M., 2015. *Error-related brain responses in high-density EEG* [Master's Thesis]. University of Freiburg, Freiburg, Germany.

Wenke, D., Atmaca, S., Holländer, A., Liepelt, R., Baess, P., Prinz, W., 2011. What is Shared in Joint Action? Issues of Co-representation, Response Conflict, and Agent Identification. *Rev. Philos. Psychol.* 2, 147–172. doi:10.1007/s13164-011-0057-0

Yeung, N., Botvinick, M.M., Cohen, J.D., 2004. The Neural Basis of Error Detection: Conflict Monitoring and the Error-Related Negativity. *Psychol. Rev.* 111, 931–959. doi:10.1037/0033-295X.111.4.931

Yordanova, J., Falkenstein, M., Hohnsbein, J., Kolev, V., 2004. Parallel systems of error processing in the brain. *NeuroImage* 22, 590–602. doi:10.1016/j.neuroimage.2004.01.040

Zander, T., Kothe, C., Jatzev, S., Luz, M., Mann, A., Welke, S., Dashuber, R., Rötting, M., 2007. Das PhyPA-BCI – Ein Brain-Computer-Interface als kognitive Schnittstelle in der Mensch-Maschine-Interaktion. Presented at the 7. *Berliner Werkstatt Mensch-Maschine-Systeme*, Berlin, Germany.